\documentclass{new_tlp}
\usepackage{url}
\usepackage{yfonts}

\newtheorem{lem}{Lemma}
\newtheorem{thm}{Theorem}

\def\ar{\leftarrow}
\def\lrar{\leftrightarrow}
\def\beq{\begin{equation}}
\def\eeq#1{\label{#1}\end{equation}}
\def\ba{\begin{array}}
\def\ea{\end{array}}
\def\clingo{{\sc clingo}}
\def\anthem{{\sc anthem}}
\def\numeral{\overline}
\def\p2f{\hbox{p2f}}
\def\no{\emph{not\/}}
\def\head{\emph{Head\/}}
\def\body{\emph{Body\/}}
\def\val#1#2{\emph{val\,}_{#1}({#2})}

\begin{document}

\title[On Program Completion]{On Program Completion, with an Application\\
  to the Sum and Product Puzzle}
\author[Vladimir Lifschitz]{Vladimir Lifschitz\\ University of Texas}
\date{}
\maketitle

\begin{abstract}
  This paper describes a generalization of Clark's completion that is
  applicable to logic programs containing arithmetic operations and
  produces syntactically simple, natural looking formulas.
  If a set of first-order axioms is equivalent to the completion of a
  program then we may be able to find standard models of these
  axioms by running an answer set solver.  As an example, we apply
  this ``reverse completion'' procedure to the Sum and Product Puzzle.
\end{abstract}

\section{Introduction}

Program completion \cite{cla78,llo84a} is a transformation that converts
logic programs into sets of first-order formulas.  The study of
completion improved our understanding of the relationship between these two
knowledge representation formalisms, and it has been used in the
design of answer set solvers \cite{lie04,lin04}.

The definition of completion has been extended to
programs with operations on integers \cite{fan20}.  That
generalized completion process produces formulas in a two-sorted first-order
language \cite[Section~5]{lif19}.
In addition to ``general'' variables, which range over
both symbolic constants and (symbols for) integers, a formula in that
language may include also variables ranging over integers only.
The need to
use a language with two sorts is explained by the fact that function symbols
in a first-order language are supposed to represent total functions, and
arithmetic operations are not defined on symbolic constants.
In answer set programming languages, applying arithmetic operations to
symbolic constants is usually handled in a different way; when a rule is
instantiated, a substituition is not used unless it is
``well-formed'' \cite[Section~3]{aspcore20}.

In this paper, the idea of a natural translation \cite{lif21} is used
to define a version of generalized completion
that is limited to relatively simple (``regular'') rules but
produces simpler, and more natural-looking,
formulas.  The modified completion operator is denoted by NCOMP, for
``natural completion.''  For example, the natural
completion of the one-rule program
\beq
\verb|even(2*X) :- X = -10..10.|
\eeq{r1}
in the input language of the answer set solver \clingo\ \cite{gringomanual}
is the sentence
\beq
\forall V (\emph{even\/}(V) \lrar \exists I (\numeral {-10}
\leq I\leq \numeral{10} \,\land\, V = \numeral 2*I)).
\eeq{c1}
Here~$V$ is a general variable,~$I$ is an integer variable, and
$\numeral {-10}$, $\numeral{10}$, $\numeral 2$ are
``numerals''---object constants representing integers.

Two theorems, stated in Section~\ref{ssec:relation} and proved in
Section~\ref{sec:proofs}, relate stable models
of a regular program to standard models of its completion
(standard in the sense that they interpret symbols related to
integers as usual in arithmetic).  These theorems
extend well-known results due to Fran\c{c}ois Fages \cite{fag91}.


If a set of first-order axioms happens to be equivalent to the completion of a
regular program then we may be able to find standard models of these
axioms by running an answer set solver.
As an example, we apply this ``reverse completion'' procedure to a
formalization of the Sum and Product Puzzle
(\url{https://en.wikipedia.org/wiki/Sum_and_Product_Puzzle}).
From the perspective of knowledge representation
and automated reasoning, that puzzle presents a challenge: express it in a
formal declarative language so that the answer can be found, or at least
verified, by an automated reasoning tool.  This has been accomplished
using first-order axioms for Kripke-style possible worlds and the
first-order theorem prover {\sc fol} \cite{mcc90a}, and also using a
modal logic of public announcements and the epistemic model checker
{\sc demo} \cite{dit05}.
More recently, Jayadev Misra
proposed a  simple first-order formalization
that does not refer to possible worlds  \cite[Section~2.8.3]{mis22}.
In Section~\ref{sec:puzzle} we show that the
answer to the puzzle can be found by applying the reverse completion
process to a variant of his axiom set and then running \clingo.

\section{Review: rules and formulas}

\subsection{Regular rules}

To simplify presentation, we do not include here some
of the programming constructs that are classified as regular in the previous
publication on natural translations
\cite{lif21}.  As in the Abstract Gringo article
\cite{geb15}, rules will be written in
abstract notation, which disregards some details related to representing
rules by strings of ASCII characters.  For example, rule~(\ref{r1}) will be
written as
\beq
\emph{even\/}(\numeral 2\times X) \ar X = \numeral {-10}\,..\,\numeral{10}.
\eeq{r1a}

We assume that
three disjoint countably infinite sets of symbols are selected:
\emph{numerals}, \emph{symbolic constants}, and \emph{(general) variables}.
We assume that a 1-1 correspondence between numerals
and integers is chosen; the numeral corresponding to an integer~$n$ is
denoted by $\numeral n$.  \emph{Precomputed terms} are numerals and
symbolic constants.  We assume that a total order on the set of precomputed
terms is chosen so that,  for all integers~$m$ and~$n$,
\begin{itemize}
\item $\numeral m < \numeral n$ iff $m<n$, and
\item
  every precomputed term~$t$ such that $\numeral m < t < \numeral n$
  is a numeral.
\end{itemize}

\emph{Regular terms} are formed from numerals and variables using
the binary function symbols $+,\ -,\ \times$.
A \emph{regular atom} is an expression of the form $p({\bf t})$, where~$p$
is a symbolic constant and $\bf t$ is a tuple of symbolic constants and
regular terms, separated by commas.  \emph{Regular comparisons}
are expressions of the forms
\begin{itemize}
\item
  $t_1\prec t_2$, where each of $t_1$, $t_2$ is a symbolic constant or a
  regular term, and~$\prec$ is one of the comparison symbols
  $=,\ \neq,\ <,\ >,\ \leq,\ \geq,$
  and
\item $t_1=t_2\,..\,t_3$, where~$t_1,\ t_2,\ t_3$ are regular terms.
\end{itemize}

A \emph{regular rule} is an expression of the form
  \beq
\head \ar \body
\eeq{rule}
where
\begin{itemize}
\item
  \head\ is either a regular atom (then~(\ref{rule}) is a
  \emph{basic rule\/}), or a regular atom in braces (then~(\ref{rule})
  is a \emph{choice rule}\/), or
  empty (then~(\ref{rule}) is a \emph{constraint}), and
\item
  \body\ is a conjunction, possibly empty, of (i)~regular atoms, possibly
  preceded by \no, and (ii)~regular comparisons.
\end{itemize}
For example,~(\ref{r1a}) is a regular rule.

A \emph{regular program} is a finite set of regular rules.  This is a
special case of Abstract Gringo programs \cite{geb15}, and stable models
of a regular program are understood in the sense of the semantic of
Abstract Gringo.  Thus stable models are sets of ground atoms that do not
contain  arithmetic operations.

\subsection{Two-sorted formulas} \label{ssec:2sorted}

A \emph{predicate symbol} is a pair $p/n$, where~$p$ is a symbolic constant
and~$n$ is a nonnegative integer. About a predicate symbol~$p/n$ we
say that it \emph{occurs} in a regular program~$\Pi$ if some atom of the
form $p(t_1,\dots,t_n)$ occurs in one of the rules of~$\Pi$.

For any regular program~$\Pi$, by $\sigma_\Pi $ we denote the two-sorted
signature with the sort \emph{general} and its
subsort \emph{integer}, which includes
\begin{itemize}
\item
  every numeral as an object constant of the sort \emph{integer},
  \item
  every symbolic constant as an object constant of the sort
  \emph{general},
\item
  the symbols $+$, $-$, $\times$ as binary function constants with the
  argument sorts \emph{integer} and the value sort \emph{integer},
\item every predicate symbol $p/n$ that occurs in~$\Pi$ as an $n$-ary
  predicate constant with the argument sorts \emph{general},
\item
  the symbols $\neq$, $<$, $>$, $\leq$, $\geq$ as binary predicate
  constants with the argument sorts \emph{general}.
\end{itemize}

A formula over $\sigma_\Pi $ that has the form $(p/n)({\bf t})$ can be
abbreviated as $p({\bf t})$.   This convention allows us to view regular
atoms occurring in~$\Pi$ as atomic formulas over~$\sigma_\Pi $.

Conjunctions of equalities and inequalities
can be abbreviated as usual in algebra; for instance, $t_1\leq t_2\leq t_3$
stands for $t_1\leq t_2\land t_2\leq t_3$.  An equality between tuples of
terms $(t_1,\dots,t_k)=(t'_1,\dots,t'_k)$ is understood as the conjunction
$t_1=t'_1\land\cdots\land t_k=t'_k$.

In this paper, integer variables are denoted by capital letters from the
middle of the alphabet ($I,\dots,N$), and general variables by letters from
the end ($U,\dots,Z$).

\section{Completion}

\subsection{Replacing variables} \label{ssec:replacing}

In the process of constructing the natural completion of a regular
program~$\Pi$, the bodies of rules of~$\Pi$ will be transformed into
formulas over~$\sigma_\Pi$.  Since general variables are not allowed in
a formula in the scope of an arithmetic operation, this process has
to involve replacing some of them by integer variables.

A \emph{critical variable} of a regular rule~$R$ is a general variable~$X$
such that at least one occurrence of~$X$ in~$R$ is in the
scope of an arithmetic operation or is part of a comparison of the form
$t_1=t_2\,..\,t_3$. For every regular rule~$R$, choose a
function~$f_R$ that maps its critical variables to pairwise distinct
integer variables.  This function~$f_R$ is extended
to other subexpressions of~$R$ as follows.  For a tuple~$\bf t$ of
symbolic constants and regular terms, $f_R({\bf t})$
is the tuple of terms over $\sigma_\Pi$ obtained from~$\bf t$ by replacing all
occurrences of every critical variable~$X$ with the integer variable~$f_R(X)$.
Applying~$f_R$ to a regular atom and to a comparison that does not contain
intervals is defined in a similar way.  The result of applying~$f_R$ to
$\no\ A$ is  defined as the
formula $\neg f_R(A)$, and the result of applying~$f_R$
to a comparison $t_1=t_2\,..\,t_3$ is $f_R(t_2)\leq f_R(t_1)\leq f_R(t_3)$.
Finally, applying~$f_R$
to the body $B_1\land B_2\land\cdots$ of~$R$ gives the formula
$f_R(B_1)\land f_R(B_2)\land\cdots$.

For instance, if~$R$ is rule~(\ref{r1a}) then the variable~$X$ is critical,
and $f_R$ maps~$X$ to some integer variable~$I$.
It transforms the term $\numeral 2\times X$ in the head into
$\numeral 2\times I$, and the body
$X = \numeral {-10}\,..\,\numeral{10}$
into
$\numeral {-10}\leq I\leq\numeral{10}$.

\subsection{Completed definitions} \label{ssec:compdefs}

Consider a regular program~$\Pi$ and a predicate symbol~$p/n$
that occurs in~$\Pi$.  The \emph{definition} of~$p/n$ in~$\Pi$
is the set of all rules of~$\Pi$ that have the form
\beq
        p({\bf t}) \ar \body
\eeq{def1}
or
\beq
        \{p({\bf t})\} \ar \body
\eeq{def2}
such that the length of the tuple~$\bf t$ is~$n$.
The \emph{completed definition} of~$p/n$ in~$\Pi$
is the sentence over $\sigma_\Pi $ constructed as follows.
Choose a tuple~$\bf V$ of~$n$ general variables that do not occur in~$\Pi$.
For every rule~$R$ in the definition~$D$ of~$p/n$ in~$\Pi$, by~$F_R$ we
denote the formula
$$
f_R(\body) \land {\bf V}=f_R({\bf t})
$$
if $R$ is~(\ref{def1}), and
$$
f_R(\body) \land {\bf V}=f_R({\bf t}) \land p({\bf V})
$$
if $R$ is~(\ref{def2}).
The completed definition of~$p/n$ in~$\Pi$ is the sentence
\beq
\forall {\bf V} \left(p({\bf V}) \lrar \bigvee_{R\in D}
  \exists {\bf U}_R F_R \right),
\eeq{compdef}
where ${\bf U}_R$ is the list of all variables occurring in $f_R(\body)$ or in
$f_R({\bf t})$.

For example, if the only rule~$R$ of the program is~(\ref{r1a}), and $p/n$
is $\emph{even\/}/1$, then~$F_R$
is
$$\numeral {-10} \leq I\leq \numeral{10} \,\land\, V = \numeral 2*I,$$
where~$I$ is $f_R(X)$.  The completed definition of $\emph{even\/}/1$
is~(\ref{c1}).

The formula obtained from the completed definition (\ref{compdef})
by replacing the global variables~$\bf V$ with fresh integer variables
will be called the \emph{arithmetic completed definition} of~$p/n$
in~$\Pi$.  For example, the arithmetic completed definition of
$\emph{even\/}/1$ in program~(\ref{r1a}) is
\beq
\forall N (\emph{even\/}(N) \lrar \exists I (\numeral {-10}
\leq I\leq \numeral{10} \,\land\, N = \numeral 2*I)).
\eeq{arcomp}
The arithmetic completed definition is
entailed by the completed definition, but not the other way around.
For example, from formula~(\ref{c1}) we can derive
$\neg\emph{even\/}(t)$ for every symbolic constant~$t$, but this conclusion
is not warranted by formula~(\ref{arcomp}).

\subsection{Natural completion}

The \emph{natural completion} NCOMP$(\Pi)$ of a regular program~$\Pi$
is the set of sentences that includes
\begin{itemize}
\item for every predicate symbol~$p/n$ occurring in~$\Pi$, its completed
  definition in~$\Pi$, and
\item for every constraint $\ar\body$ in~$\Pi$,
  the universal closure of the formula
$$\neg f_{\ar Body}(\body).$$
\end{itemize}

Consider, for example,
the program that consists of rule~(\ref{r1a}), the choice rule
\beq
\{\emph{foo\/}(X)\} \ar \emph{even\/}(X)
\eeq{r2}
and the constraint
\beq
\ar \no\ \emph{foo\/}(\numeral 0).
\eeq{r3}
Its natural completion consists of the completed definition~(\ref{c1})
of~$\emph{even\/}/2$, the
completed definition of~$\emph{foo\/}/1$
$$
\forall V(\emph{foo\/}(V) \lrar \exists X(\emph{even\/}(X) \land X=V \land \emph{foo\/}(V))),
$$
which can be rewritten\footnote{When we talk about equivalent transformations
  of a completed definition, equivalence is understood in the sense of
  classical first-order logic.} as
$$\forall V(\emph{foo\/}(V) \to \emph{even\/}(V)),$$
and the sentence $\neg\neg \emph{foo\/}(\numeral 0)$,
which is equivalent to $\emph{foo\/}(\numeral 0)$.

\subsection{Relation to stable models} \label{ssec:relation}

The \emph{Herbrand base} of a regular program~$\Pi$ is the set of
all regular atoms $p(t_1,\dots,t_n)$ such that $p/n$ occurs in~$\Pi$ and
$t_1,\dots,t_n$ are precomputed terms.  If~$S$ is a subset of
the Herbrand base of~$\Pi$ then $S^\uparrow$ is the interpretation of
the signature $\sigma_\Pi$ defined as follows:
\begin{itemize}
\item[(i)]
  the universe of the sort \emph{general} in $S^\uparrow$ is the set of all
  precomputed terms;
\item[(ii)]
  the universe of the sort \emph{integer} in $S^\uparrow$ is the set of all
  numerals;
\item[(iii)]
  for every precomputed term~$t$, $S^\uparrow(t)=t$;
\item[(iv)]
  for every pair~$m$,~$n$ of integers,
 $S^\uparrow(\numeral m+\numeral n)=\numeral{m+n}$,
  and similarly for subtraction and multiplication;
\item[(v)]
  for every pair~$t_1,~t_2$ of precomputed terms,
  $S^\uparrow$ satisfies $t_1<t_2$ iff the relation~$<$ holds for
  the pair~$t_1$,~$t_2$, and similarly for the other comparison symbols.
\end{itemize}

\begin{thm}\label{thm1}
  For any regular program~$\Pi$ and any subset~$S$ of its Herbrand base,
  if~$S$ is a stable model of~$\Pi$ then~$S^\uparrow$
  satisfies NCOMP$(\Pi)$.
\end{thm}

The \emph{positive predicate dependency graph} of a regular program~$\Pi$ is
the directed graph defined as follows.  Its vertices are the predicate
symbols $p/n$ occurring in~$\Pi$.  It has an edge from $p/n$ to $q/m$ if~$\Pi$
has a rule~(\ref{rule}) such that
\begin{itemize}
\item \head\ has the form $p(t_1,\dots,t_n)$ or $\{p(t_1,\dots,t_n)\}$, and
\item one of the conjunctive terms of \body\ has the form
  $q(t_1,\dots,t_m)$.
\end{itemize}
A regular program~$\Pi$ is \emph{tight} if its positive predicate dependency
graph is acyclic.

For example, the positive predicate dependency graph of
program~(\ref{r1a}),~(\ref{r2}),~(\ref{r3}) has one edge, from
$\emph{foo\/}/1$ to $\emph{even\/}/1$.  This program is tight.

\begin{thm}\label{thm2}
  For any tight regular program~$\Pi$ and any subset~$S$ of its
  Herbrand base,~$S$ is a stable model of~$\Pi$ iff~$S^\uparrow$
  satisfies NCOMP$(\Pi)$.
\end{thm}

\section{The puzzle}\label{sec:puzzle}

\emph{Two mathematicians, S and P, talk about two integers, $M$ and $N$.
S knows the sum $M+N$, and P knows the product $M\times N$. Both S and P
know also that the integers are greater than 1; that their sum is not
greater than 100; and that $N$ is greater than $M$.
The following conversation occurs:
\begin{enumerate}
\item
S says: P does not know $M$ and $N$.
\item
P says: Now I know $M$ and $N$.
\item
S says: Now I also know $M$ and $N$.
\end{enumerate}
What are $M$ and $N$?
}

\subsection{First-order axioms}

Jayadev Misra's approach to translating this puzzle into a
first-order language \cite[Section~2.8.3]{mis22} involves the use of binary
predicates~$b_0,\dots,b_3$.  The formula $b_0(M,N)$ expresses that
before the beginning of the conversation the pair~$M$, $N$ was
considered a possible solution.  This can be expressed by the formula
\beq
b_0(M,N) \lrar 1 < M < N \land M+N \leq 100.
\eeq{ax1}
The formula $b_1(M,N)$ expresses that~$M$, $N$ was considered a
possible solution at step 1, that is, after hearing the words
``P does not know $M$ and $N$''; and so forth.

There are several ways to write axioms for $b_1$, $b_2$ and $b_3$.  One
possibility is described below.

We say that an integer~$I$ is \emph{puzzling at time 0} if
is there is more than one way
to represent it as the product of two numbers $J$, $K$ satisfying
$b_0(J,K)$:
\beq
\ba r
\!\!\emph{puzzling\/}_0(I)\lrar \exists J_1K_1J_2K_2(b_0(J_1,K_1)
\land b_0(J_2,K_2)\hskip 2.25cm \\
\land\, I=J_1\times K_1 =J_2\times K_2\land J_1\neq J_2).
\ea
\eeq{ax2}
We say that an integer~$I$ is \emph{possibly easy} if it can be represented
as the sum of two numbers~$J$,~$K$ satisfying $b_0(J,K)$ such that
$J\times K$ is not puzzling at time~0:
\beq
\emph{possibly\_easy\/}(I) \lrar \exists JK(b_0(J,K)\land I = J+K
\land\neg\emph{puzzling\/}_0(J\times K)).
\eeq{ax3}
Then the assumption
$$\hbox{``at Step 1, S knows that P does not know $M$ and $N$''}$$
can be expressed by the axiom
\beq
b_1(M,N) \lrar b_0(M,N)\land\neg\emph{possibly\_easy\/}(M+N).
\eeq{ax4}
We say that an integer~$I$ is \emph{puzzling at time 1}
is there is more than one way
to represent it as the product of two numbers $J$, $K$ satisfying
$b_1(J,K)$:
\beq
\ba r
\!\!\emph{puzzling\/}_1(I)\lrar \exists J_1K_1J_2K_2(b_1(J_1,K_1)
\land b_1(J_2,K_2)\hskip 2.25cm \\
\land\, I=J_1\times K_1 =J_2\times K_2\land J_1\neq J_2).
\ea
\eeq{ax5}
The assumption
$$\hbox{``at Step 2, P knows $M$ and $N$''}$$
can be expressed by the axiom
\beq
b_2(M,N) \lrar b_1(M,N)\land\neg\emph{puzzling\/}_1(M\times N).
\eeq{ax6}
We say that an integer~$I$ is \emph{puzzling at time 2}
is there is more than one way
to represent it as the sum of two numbers $J$, $K$ satisfying
$b_2(J,K)$:
\beq
\ba r
\!\!\emph{puzzling\/}_2(I)\lrar \exists J_1K_1J_2K_2(b_2(J_1,K_1)
\land b_2(J_2,K_2)\hskip 2.25cm \\
\land\, I=J_1+K_1 =J_2+K_2\land J_1\neq J_2).
\ea
\eeq{ax7}
Finally, the assumption
$$\hbox{``at Step 3, S knows $M$ and $N$''}$$
can be expressed by the axiom
\beq
b_3(M,N) \lrar b_2(M,N)\land\neg\emph{puzzling\/}_2(M+N).
\eeq{ax8}

Since axioms (\ref{ax1})--(\ref{ax8}) form a chain of explicit definitions,
the predicates represented by the symbols
\beq
b_0/2,\ \dots,\ b_3/2,\ \emph{puzzling\/}_0/1,\ \dots,\ 
\emph{puzzling\/}_2/1,\ \emph{possibly\_easy\/}/1
\eeq{preds}
are uniquely defined,
assuming that variables range over the integers and that the symbols
$$+\quad\times\quad<\quad\leq$$
are interpreted in the standard way.  To solve the Sum and Product Puzzle,
we will calculate the extents of these predicates.

This will be accomplished by running \clingo\ on the ``reverse completion''
of axioms~(\ref{ax1})--(\ref{ax8}). 

\subsection{Reverse completion}

Consider the regular program
\beq\ba{rcl}
b_0(\emph{XM\/},\emph{XN\/}) &\ar& \numeral 1 < \emph{XM\/}\,\land\, \emph{XM\/} < \emph{XN\/}\,\land\, \emph{XM\/}+\emph{XN\/} \leq \numeral{100},\\

\emph{puzzling\/}_0(\emph{XI}) &\ar& b_0(\emph{XJ}_1,\emph{XK}_1)\,\land\, b_0(\emph{XJ}_2,\emph{XK}_2)\,\land\, \emph{XI} = \emph{XJ}_1 \times \emph{XK}_1\\
&&\hskip 1cm \land\;  \emph{XJ}_1 \times \emph{XK}_1=\emph{XJ}_2 \times \emph{XK}_2\,\land\, \emph{XJ}_1 \neq  \emph{XJ}_2,\\

\emph{possibly\_easy\/}(\emph{XI}) &\ar& b_0(\emph{XJ},\emph{XK})\,\land\, \emph{XI} = \emph{XJ}+\emph{XK}\\
&&\land\; \no\ \emph{puzzling\/}_0(\emph{XJ} \times \emph{XK}),\\
b_1(\emph{XM\/},\emph{XN\/})&\ar& b_0(\emph{XM\/},\emph{XN\/})\,\land\, \no\ \emph{possibly\_easy\/}(\emph{XM\/}+\emph{XN\/}),\\

\emph{puzzling\/}_1(\emph{XI}) &\ar& b_1(\emph{XJ}_1,\emph{XK}_1)\,\land\, b_1(\emph{XJ}_2,\emph{XK}_2)\,\land\, \emph{XI} = \emph{XJ}_1 \times \emph{XK}_1\\
&&\hskip 1cm\land\;  \emph{XJ}_1 \times \emph{XK}_1=\emph{XJ}_2 \times \emph{XK}_2\,\land\, \emph{XJ}_1 \neq  \emph{XJ}_2,\\

b_2(\emph{XM\/},\emph{XN\/}) &\ar& b_1(\emph{XM\/},\emph{XN\/})\,\land\, \no\ \emph{puzzling\/}_1(\emph{XM\/} \times \emph{XN\/}),\\

\emph{puzzling\/}_2(\emph{XI}) &\ar& b_2(\emph{XJ}_1,\emph{XK}_1)\,\land\, b_2(\emph{XJ}_2,\emph{XK}_2)\,\land\, \emph{XI} = \emph{XJ}_1+\emph{XK}_1\\
&&\hskip 1cm\land\;  \emph{XJ}_1+\emph{XK}_1=\emph{XJ}_2+\emph{XK}_2\,\land\,\  \emph{XJ}_1 \neq  \emph{XJ}_2,\\

b_3(\emph{XM\/},\emph{XN\/}) &\ar& b_2(\emph{XM\/},\emph{XN\/})\,\land\,\no\ \emph{puzzling\/}_2(\emph{XM\/}+\emph{XN\/}).
\ea
\eeq{sp}
These rules are obtained from equivalences (\ref{ax1})--(\ref{ax8}) by
\begin{itemize}
\item replacing the equivalence signs $\lrar$ by left arrows,
\item dropping existential quantifiers,
\item replacing integer variables by general variables, and
\item replacing~$\neg$ by \no.
\end{itemize}

The natural completion of program~(\ref{sp}) looks very similar to
axiom set (\ref{ax1})--(\ref{ax8}).  There is a difference though: the
former consists of formulas over the two-sorted signature described in
Section~\ref{ssec:2sorted}, and the axioms formalizing the Sum and
Product puzzle are one-sorted; there are no
general variables in them.  Consider then the 
\emph{arithmetic} completed definitions of predicate symbols~(\ref{preds})
(see Section~\ref{ssec:compdefs}).
Those are one-sorted formulas, and they are
equivalent to the universal closures of the
corresponding axioms.  For example, the completed definition of the
predicate symbol~$b_0/2$ in program~(\ref{sp}) is
$$\ba l
\forall \emph{XM\/}\,\emph{XN\/}(b_0(\emph{XM\/},\emph{XN\/}) \lrar\\
\hskip 1.8cm
\exists M N(\numeral 1 < M\land M < N\land M+N \leq \numeral{100}
\land \emph{XM\/}=M\land \emph{XN\/}=N)),
\ea$$
and the arithmetic completed definition of this symbol is the one-sorted
formula
$$\ba l
\forall M_1N_1(b_0(M_1,N_1) \lrar\\
\hskip 1.8cm
\exists M N(\numeral 1 < M\land M < N\land M+N \leq \numeral{100}
\land M_1=M\land N_1=N)).
\ea$$
This formula is equivalent to the universal closure of axiom~(\ref{ax1}).
Similarly, the arithmetic completed definition
of~$\emph{puzzling\/}_0/2$ is equivalent to the universal closure
of axiom~(\ref{ax2}), and so forth.

\subsection{Calculating the answer}

By running \clingo\ we can determine that program~(\ref{sp}) has a
unique stable model~$S$.  By Theorem~\ref{thm1}, the
interpretation~$S^\uparrow$ satisfies the completed definitions of
symbols~(\ref{preds}).  It follows that~$S^\uparrow$ satisfies the arithmetic
completed definitions of these symbols, which are equivalent to
axioms~(\ref{ax1})--(\ref{ax8}).  In other words,~$S$ describes
the extents of the predicates that we want to calculate.
Since the only atom in~$S$ that begins with~$b_3$ is
$b_3(\numeral 4,\numeral{13})$, the answer to the puzzle is
$$M=4,\ N=13.$$

To perform this calculation, we used
version~5.6.0 of \clingo.  Earlier versions do
not accept the first rule of the program as safe unless the expression
$$\numeral 1 < M\land M < N$$
in the body is rewritten in
the interval notation: $M=\numeral 2\,..\,N-\numeral 1$.

\section{Proofs}\label{sec:proofs}

Proofs of Theorems~\ref{thm1} and~\ref{thm2} are based on similar results
from an earlier publication \cite{fan20}, and we begin by reviewing them for
the special case of regular programs.

\subsection{Review: Completion according to Fandinno et al.}
                                           \label{ssec:fea}

For any regular term~$t$, the formula $\val tZ$, where~$Z$ is a general
variable that does not occur in~$t$, is defined recursively:
\begin{itemize}
\item
if $t$ is a numeral or a variable then $\val tZ$ is $Z=t$;
\item
if $t$ is $t_1 + t_2$ then $\val tZ$ is
$$ 
\exists I J  (Z=I + J \land \val{t_1}I \land \val{t_2}J),
$$ 
and similarly for $t_1-t_2$ and $t_1\times t_2$.
\end{itemize}
If~$t$ is a symbolic constant then $\val tZ$ stands for $Z=t$.
If $t_1,\dots,t_k$ is a tuple of symbolic constants and regular terms, and
$Z_1,\dots,Z_k$ are pairwise distinct general variables that do not occur in
$t_1,\dots,t_k$, then $\val{t_1,\dots,t_k}{Z_1,\dots,Z_k}$ stands for
$\val{t_1}{Z_1} \land \cdots \land \val{t_k}{Z_k}$.
If $t$ is $t_1\,..\,t_2$ then $\val tZ$ stands for
$$\exists I J K (\val{t_1}I \land \val{t_2}J \land
    I\leq K \leq J \land Z=K).$$

The translation $\tau^B$ transforms expressions in the body of a regular rule
into formulas as follows:
\begin{itemize}
\item
$\tau^B(p({\bf t}))$ is
$\exists {\bf Z}(\val{\bf t}{\bf Z} \land p({\bf Z}))$,
where $\bf Z$ is a tuple of distinct program variables that do not occur
in~$\bf t$;
\item
$\tau^B(\no\ p({\bf t}))$ is
$\exists {\bf Z}(\val{\bf_t}{\bf Z} \land \neg p({\bf Z}))$;
\item
$\tau^B(t_1\prec t_2)$ is
$\exists Z_1 Z_2 (\val{t_1,t_2}{Z_1,Z_2} \land Z_1\prec Z_2)$;
\item
$\tau^B(t_1=t_2\,..\,t_3)$ is
$\exists Z_1 Z_2 (\val{t_1}{Z_1}\land \val{t_2\,..\,t_3}{Z_2}\land Z_1=Z_2)$;
\item
  $\tau^B(B_1\land B_2\land\cdots)$ is
  $\tau^B(B_1)\land\tau^B(B_2)\land\cdots$.
\end{itemize}

The \emph{completed definition of~$p/n$ in~$\Pi$
in the sense of Fandinno et al.}~is the sentence over $\sigma_\Pi$
constructed as follows.
Choose a tuple~$\bf V$ of~$n$ general variables that do not occur in~$\Pi$.
For every rule~$R$ in the definition~$D$ of~$p/n$ in~$\Pi$, by~$F'_R$ we
denote the formula
$$
\tau^B(\body) \land \val{\bf t}{\bf V}
$$
if $R$ is~(\ref{def1}), and
$$
\tau^B(\body) \land \val{\bf t}{\bf V} \land p({\bf V})
$$
if $R$ is~(\ref{def2}).
The completed definition of~$p/n$ in~$\Pi$ according to Fandinno \emph{et
  al.}~is the sentence
$$\forall {\bf V} \left(p({\bf V}) \lrar \bigvee_{R\in D}
 \exists {\bf U}'_R F'_R \right),$$
where ${\bf U}'_R$ is the list of all variables occurring in~$R$.

For example, the completed definition of one-rule program~(\ref{r1a}) is
\beq
\forall V \left(\emph{even\/}(V) \lrar
  \exists X \left(\exists Z_1Z_2\left(Z_1=X\land
    \val{\numeral {-10}\,..\,\numeral{10}}{Z_2}
    \land Z_1=Z_2\right)
  \land \val{\numeral 2\times X}V\right)\right),
\eeq{longcd}
where $\val{\numeral {-10}\,..\,\numeral{10}}{Z_2}$ stands for
$$\exists I J K (I=\numeral{-10} \land J=\numeral{10} \land
    I\leq K \leq J \land Z_2=K)$$
and $\val{\numeral 2\times X}V$ stands for
$$\exists I J  (V=I \times J \land I =\numeral 2 \land J = X).$$
    
By COMP$(\Pi)$ we denote the set of sentences that includes
\begin{itemize}
\item for every predicate symbol~$p/n$ occurring in~$\Pi$, its completed
  definition in~$\Pi$ in the sense of Fandinno et al, and
\item for every constraint $\ar\body$ in~$\Pi$,
  the universal closure of the formula
  $\neg \tau^B(\body)$.
\end{itemize}

\begin{lem}\label{lem1}
  For any regular program~$\Pi$ and any subset~$S$ of its Herbrand base,
  if~$S$ is a stable model of~$\Pi$ then~$S^\uparrow$
  satisfies COMP$(\Pi)$ \cite[Theorem 1]{fan20}.
\end{lem}

\begin{lem}\label{lem2}
  For any tight regular program~$\Pi$ and any subset~$S$ of its Herbrand base,
  $S$ is a stable model of~$\Pi$ iff~$S^\uparrow$ satisfies
  COMP$(\Pi)$ \cite[Theorem 2]{fan20}.
\end{lem}

\subsection{Main Lemma}

Theorems~\ref{thm1} and~\ref{thm2} follow from Lemmas~\ref{lem1} and~\ref{lem2}
in view of the following fact, proved below:

\medskip\noindent\emph{Main Lemma}\\
For any regular program~$\Pi$,
  the formula NCOMP$(\Pi)$ is equivalent to COMP$(\Pi)$ in
  classical predicate calculus with equality.
\medskip

In the statements of Lemmas~\ref{lem3}--\ref{lem7}, $R$ is a regular rule,
$\bf X$ is the list of its critical variables, and $\bf I$ is $f_R({\bf X})$.

\begin{lem}\label{lem3}
If~$\bf t$ is a list of symbolic constants and regular terms that
occur in~$R$, and $\bf Z$ is a list of pairwise distinct general
variables that do not occur in~$R$, then the formula
\beq
{\bf I}={\bf X} \to
\forall {\bf Z}(val_{\bf t}({\bf Z}) \lrar {\bf Z}=f_R({\bf t}))
\eeq{lem3f}
  is logically valid \cite[Lemma~1(i)]{lif21}.
\end{lem}
    
\begin{lem}\label{lem4}
For any regular atom~$p({\bf t})$ occurring in~$R$, the formulas
$$\ba l
{\bf I}={\bf X} \to (f_R(p({\bf t})) \lrar \tau^B(p({\bf t}))),\\
{\bf I}={\bf X} \to (f_R(not\ p({\bf t})) \lrar \tau^B(not\ p({\bf t})))
\ea$$
are logically valid \cite[Lemma~1(iii,iv)]{lif21}.
\end{lem}

\begin{lem}\label{lem5}
For any comparison $t_1\prec t_2$ occurring in~$R$, the formula
$${\bf I}={\bf X} \to (f_R(t_1\prec t_2) \lrar \tau^B(t_1\prec t_2))$$
is logically valid \cite[Lemma~2]{lif21}.
\end{lem}

\begin{lem}\label{lem6}
For any comparison $t_1 = t_2\,..\,t_3$ occurring in~$R$, the formula
$${\bf I}={\bf X} \to (f_R(t_1 = t_2\,..\,t_3)
\lrar \tau^B(t_1 = t_2\,..\,t_3))$$
is logically valid \cite[Lemma~4]{lif21}.
\end{lem}

From Lemmas~\ref{lem4}--\ref{lem6} we conclude:

\begin{lem}\label{lem7}
The formula
\beq{\bf I}={\bf X} \to (f_R(Body)\lrar \tau^B(Body)),\eeq{l7}
where \emph{Body} is the body of~$R$, is logically valid.
\end{lem}

\begin{lem}\label{lem8}
Let~$R$ be a regular rule with the body $B_1\land B_2\land \cdots$.
\begin{itemize}
\item[(a)]
  For every variable~$X$ that occurs in~$B_i$ in the scope of an
  arithmetic operation, the formula
  \beq
  \tau^B(B_i)\to\exists I(I=X)
  \eeq{l8}
  is logically valid.
\item[(b)]
  If~$B_i$ is a comparison of the form $t_1=t_2\,..\,t_3$ then
  formula~(\ref{l8}) is logically valid for every variable~$X$ that
  occurs in~$B_i$.
\end{itemize}
\cite[Lemmas~7 and~9]{lif21}.
\end{lem}

From Lemma~\ref{lem8} we conclude:

\begin{lem}\label{lem9}
  Let~$R$ be a regular rule, and let~$\bf X$ be the list of its critical
  variables.  The formula
  \beq
  \tau^B(Body) \to \exists {\bf I}({\bf I}={\bf X}),
  \eeq{l9}
  where \emph{Body} is the body of~$R$ and $\bf I$ is $f_R({\bf X})$,
  is logically valid.
\end{lem}

In the following lemma,~$R$ is a regular rule of form~(\ref{def1})
or~(\ref{def2}), $F_R$ and ${\bf U}_R$ are as defined in
Section~\ref{ssec:compdefs}, and $F'_R$ and ${\bf U}'_R$ are as defined in
Section~\ref{ssec:fea}.

\begin{lem}\label{lem10}
The formula $\exists {\bf U}'_R F'_R$ is equivalent to $\exists {\bf U}_R F_R$.
\end{lem}

\noindent{\bf Proof}$\,$ Let~$\bf X$ be the list of all critical variables
of~$R$, and let~$\bf I$ be $f_R({\bf X})$.
It is sufficient to prove the equivalence
$$
\exists {\bf I}\, F'_R\lrar \exists {\bf X}\, F_R.
$$
We consider the case when~$R$ is~(\ref{def1}), so that $F_R$ is
$$
\tau^B(\body) \land \val{\bf t}{\bf V}
$$
and $F'_R$ is
$$
f_R(\body) \land {\bf V}=f_R({\bf t}).
$$
The case of rule~(\ref{def2}) is similar.

Left-to-right: assume $\exists {\bf I}(f_R(\body) \land {\bf V}=f_R({\bf t}))$.
Since $\forall{\bf I}\exists{\bf X}({\bf I}={\bf X})$,
we can conclude that
$$\exists {\bf I}
(\exists {\bf X}({\bf I}={\bf X}) \land f_R(\body)\land {\bf V}=f_R({\bf t})).
$$
It follows that
$$\exists {\bf IX}
({\bf I}={\bf X} \land f_R(\body)\land {\bf V}=f_R({\bf t})),$$
because the critical variables~$\bf X$ do not occur in the formula
$f_R(\body)\land {\bf V}=f_R({\bf t})$. Then
$\exists {\bf IX}(\tau^B(\body)\land \val{\bf t}{\bf V})$
follows using the iniversal closures of~(\ref{lem3f}) and~(\ref{l7}).
Since the integer variables $\bf I$ do not occur in
$\tau^B(\body)\land \val{\bf t}{\bf V}$, the quantifiers binding~$\bf I$
can be dropped.

Right-to-left: assume
$\exists {\bf X} (\tau^B(\body)\land \val{\bf t}{\bf V})$.
We can conclude, using the universal closure of~(\ref{l9}), that
$$
\exists {\bf X}(\exists {\bf I}({\bf I}={\bf X})
\land \tau^B(\body)\land \val{\bf t}{\bf V}).
$$
It follows that
$$
\exists {\bf IX}({\bf I}={\bf X}\land \tau^B(\body)\land \val{\bf t}{\bf V}),
$$
because the integer variables~$\bf I$ do not occur in the formula
$\tau^B(\body)\land \val{\bf t}{\bf V})$.
Then $\exists {\bf IX}(f_R(\body)\land {\bf V}=f_R({\bf t}))$ follows
using the iniversal closures of~(\ref{lem3f}) and~(\ref{l7}).
Since the critical variables $\bf X$ do not occur in
$f_R(\body)\land {\bf V}=f_R({\bf t})$,
the quantifiers binding~$\bf X$ can be dropped.

\begin{lem}\label{lem11}
  If~$R$ is a regular constraint $\ar Body$ then the universal closures of
  the formulas $\neg f_R(Body)$ and $\neg \tau^B(Body)$ are equivalent to
  each other.
\end{lem}

\noindent{\bf Proof}$\,$ Let~$\bf X$ be the list of all critical variables
of~$R$, and let~$\bf I$ be $f_R({\bf X})$.
It is sufficient to prove the equivalence
$$
\exists {\bf I}\, f_R(\body) \lrar \exists {\bf X}\, \tau^B(\body),
$$
because it entails
$$\forall {\bf I}\neg f_R(\body) \lrar \forall {\bf X}\neg \tau^B(\body)$$
and consequently entails also the equivalence between the universal closures of
$\neg f_R(Body)$ and $\neg \tau^B(Body)$.

Left-to-right:
assume $\exists {\bf I}\, f_R(\body)$.
Since $\forall{\bf I}\exists{\bf X}({\bf I}={\bf X})$,
we can conclude that
$$\exists {\bf IX}({\bf I}={\bf X} \land f_R(\body)),$$
because the critical variables~$\bf X$ do not occur in $f_R(\body)$.
Then $\exists {\bf IX}\, \tau^B(\body)$ follows
using the iniversal closure
of~(\ref{l7}).  Since the integer variables $\bf I$ do not occur in
$\tau^B(\body)$, the quantifiers binding~$\bf I$ can be dropped.

Right-to-left: assume $\exists {\bf X}\, \tau^B(\body)$.
We can conclude, using the universal closure of~(\ref{l9}), that
$$\exists {\bf IX}({\bf I}={\bf X}\land \tau^B(\body)),$$
because the integer variables~$\bf I$ do not occur in $\tau^B(\body)$.
Then $\exists {\bf IX}\, f_R(\body)$ follows using the iniversal closure
of~(\ref{l7}).  Since the critical variables $\bf X$ do not occur in
$f_R(\body)$, the quantifiers binding~$\bf X$ can be dropped.

\medskip
Main Lemma follows from Lemmas~\ref{lem10} and~\ref{lem11}.

\section{Discussion}

In the presence of arithmetic operations, the completed definition
in the sense of Fandinno et al.~is often longer and syntactically more complex
than the ``natural'' completed
definition introduced in Section~\ref{ssec:compdefs}; compare, for instance,
formula~(\ref{longcd}) with~(\ref{c1}).  On the other hand, the approach of
Fandinno et al.~is applicable to some types of rules that are accepted
by \clingo\ but are not regular,
such as
\begin{verbatim}
  p(1..8,1..8).
  p(2*(1..8)).
\end{verbatim}
These two rules can be easily rewritten as regular rules:
\begin{verbatim}
  p(X,Y) :- X = 1..8, Y = 1..8.
  p(2*X) :- X = 1..8.
\end{verbatim}
Regularizing the rule
\begin{verbatim}
  q(X/5) :- p(X).
\end{verbatim}
in a similar way gives the rule
\begin{verbatim}
  q(Y) :- p(X), X = 5*Y+Z, Z = 0..4.
\end{verbatim}
that the current version of \clingo\ considers unsafe.

The translation COMP is used in the design of the proof assistant
\anthem\ \cite{fan20}, and our Main Lemma shows that NCOMP
can be employed in the same way.  In the process of interacting with \anthem,
the user often has to read and modify completion formulas.  A
version of \anthem\ that implements
natural completion would make this work easier.

It would be interesting to extend the definition of NCOMP to programs
containing conditional literals \cite[Section~3.1.11]{gringomanual}.  Such an
extension would make the reverse completion process applicable to some
formulas that are more complex syntactically than the current version.  For
instance, it may be able to handle the formula
$$\ba l
b_1(M,N)\, \lrar\\
\qquad b_0(M,N)\land\neg\exists JK(b_0(J,K)\land M+N = J+K
\land\neg\emph{puzzling\/}_0(J\times K)),
\ea$$
which can replace axioms~(\ref{ax3}),~(\ref{ax4}) in the first-order
  formalization of the Sum and Product Puzzle.

\section*{Acknowledgements}

Many thanks to Jorge Fandinno, Michael Gelfond, Yuliya Lierler,
Jayadev Misra and Nathan Temple for discussions related to
the topic of this paper, and to the anonymous referees for advice on
improving the previous version.

\bibliographystyle{acmtrans}
\bibliography{bib}
\end{document}